\documentclass[onecolumn,prb,groupedaddress,preprintnumbers,amsmath,amssymb, floatfix]{revtex4}
\usepackage{amsmath}
\usepackage{graphicx}
\usepackage{dcolumn}
\begin{document}
\title{Thermoresponsive Colloidal Molecules}

\author{
Martin Hoffmann,\textsuperscript{\textit{a}} Miriam Siebenb\"urger,\textsuperscript{\textit{b}} Ludger Harnau,\textsuperscript{\textit{cd}} Markus Hund,\textsuperscript{\textit{e}} Christoph Hanske,\textsuperscript{\textit{e}} Yan Lu,\textsuperscript{\textit{b}} Claudia S. Wagner,\textsuperscript{\textit{a}} Markus Drechsler,\textsuperscript{\textit{f}} Matthias Ballauff\textsuperscript{\textit{*b}}}
\affiliation{
\small\textit{\begin{itemize}
\item[\textsuperscript{a}] Physikalische Chemie I, Universi\"at Bayreuth, 
Universit\"atsstr. 30, 95440 Bayreuth, Germany
\item[\textsuperscript{b}]F-I2 Soft Matter and Functional Materials, 
Helmholtz-Zentrum Berlin, Glienicker Str. 100, 14109 Berlin, Germany
\item[\textsuperscript{c}] Max-Planck-Institut f\"ur Metallforschung, 
Heisenbergstr. 3, D-70569, Stuttgart, Germany
\item[\textsuperscript{d}] Institut f\"ur Theoretische und Theoretische und 
Angewandte Physik, Universit\"at Stuttgart, Pfaffenwaldring 57, D-70569 Stuttgart, Germany
\item[\textsuperscript{e}] Physikalische Chemie II, Universit\"at Bayeuth, 
Universit\"atsstr. 30, 95440 Bayreuth, Germany
\item[\textsuperscript{f}] Macromolecular Chemistry II, University of Bayreuth, 95440 Bayreuth, Germany
\end{itemize}}
	 }

\begin{abstract}
We fabricated thermoresponsive colloidal molecules of ca. 250 nm
size. Electron- and scanning force microscopy reveal the dumbbell-shaped
morphology. The temperature dependence of the size and
aspect ratio (ca. 1.4 to 1.6) is analyzed by depolarized dynamic light
scattering and found to be in good agreement with microscopic
evidence.
\end{abstract}
\maketitle

Self-assembly of thermoresponsive building blocks is a promising technique to fabricate new materials with tunable properties.\textsuperscript{1} Due to their potential applications as biosensors,\textsuperscript{2} delivery systems for biomolecules\textsuperscript{3} or photonic devices,\textsuperscript{4} there is a considerable interest in this fascinating class of material. The most studied example, poly-(\textit{N}-isopropylacrylamide)
(PNIPA) undergoes a volume phase transition in water above its lower critical solution temperature (LCST) at ca. 32 $^{\rm o}$C. Spherical microgels with a solid core and a thermoresponsive shell can be used as colloidal atoms in soft matter physics to mimic transitions from the liquid to the crystalline phase including nonequilibrium phenomena like the glass transition in concentrated
suspensions.\textsuperscript{5}\\
However, despite recent advances with PNIPA-coated (spherical) core particles made of silica or gold\textsuperscript{6} or raspberry-like stimuli-responsive
polymer particles,\textsuperscript{7} no attempts have been made for different core geometries so far. Colloidal clusters\textsuperscript{8} and anisotropic nanoparticles\textsuperscript{9} with shapes resembling space-filling models of simple molecules are expected to show complex behavior like low molecular weight compounds.\textsuperscript{10} Thus, extending the analogy between atoms and colloidal spheres, such colloids can be regarded as ``colloidal molecules'', a term introduced by van Blaaderen.\textsuperscript{11} However, the geometry of these particles cannot be changed without chemical modification once synthesized.\\
In this communication, we describe a strategy that overcomes this limitation by extending our previous work on spherical particles to
dumbbell-shaped core-shell microgel particles (DMPs).We generate particles with variable morphology by changing the temperature.
Field emission scanning electron microscopy (FESEM), cryogenic-transmission electron microscopy (cryo-TEM) and scanning force microscopy (SFM) are used to obtain real space information about the particle morphology. The diffusion coefficients for translational-
($D^T$) and rotational motion ($D^R$) are determined by a combination of
polarized (DLS) and depolarized dynamic light scattering (DDLS)\textsuperscript{12} in the highly diluted regime. Modelling of the diffusion coefficients with the hydrodynamic shell-model gives access to the particle size parameters and the thickness of the PNIPA layer. In this model the particle surface is regarded as a shell of small, non-overlapping and spherical friction elements.\textsuperscript{13} It will be shown that the growth of the crosslinked PNIPA shell is not influenced by the core particle geometry. The findings present a significant step towards a general understanding of ``colloidal molecules'', as in principle the strategy is not limited to a certain particle configuration. Because the synthesis facilitates yields in the gram-scale, the DMPs are a versatile model system to investigate the fluid-solid transitions of concentrated dispersions of both thermoresponsive and anisotropic colloids.\textsuperscript{14}\\
We prepared aqueous suspensions of DMP particles with a dumbbell-shaped polymer core 
and a thermoresponsive PNIPA-shell crosslinked by 5 mol$\%$ \textit{N,N}'-methylenebisacrylamide (BIS). The polymer core particles were synthesized by seeded growth emulsion polymerization using PMMA seeds and styrene monomer added under starved conditions. \textsuperscript{12} A relatively high surface tension between the PMMA surface and styrene-monomer as well as the low reaction temperature (60 $^{\rm o}$C) and the low concentration of styrene in the water phase (kinetic control) favor the dumbbell morphology (Fig. 1a). After purification by repeated ultracentrifugation, the latex exhibits a red iridescence.\\
As opposed to previous studies on spherical core-shell microgels,\textsuperscript{15} NIPA was not copolymerized with styrene during the formation of the core particles as NIPA is expected to influence the resulting particle morphology. For the formation of the crosslinked thermoresponsive shell, we copolymerized NIPA and BIS ($95/5$ mol$\%$) in the presence of the PMMA/PS core particles. Due to initiator fragments near the particle surface, the microgels exhibit a zeta-potential of $-30.7$ mV in water and $-2.8$ mV in 5 mM KCl at $\text{25 $^{\rm o}$C}$. If the microgels are deposited on a weak negatively charged Si wafer, the electrostatic repulsions between the microgels among themselves and with the substrate lead to the 2-dimensional pattern as shown in Fig. 1b (FESEM micrograph). The spreading of the collapsed PNIPA shell on the Si surface is more homogeneous than for negatively charged spherical core-shell microgels as reported recently.\textsuperscript{16} Notably, at room temperature the distance between the microgels in the dried state (ca. $250\pm50$ nm) is larger than twice the thickness of the PNIPA layer in the wet state (ca. $100\pm10$ nm) as will be shown by cryo-TEM or DDLS/DLS measurements (see below). This suggests an electrostatic repulsion of the microgels leading to a superlattice when water is evaporated. After centrifugation of diluted DMP solutions, the particles show greenish to bluish iridescence depending on the angle of incident light and observation.\\
To investigate the particle morphology \textit{in situ}, cryo-TEM measurements\textsuperscript{15\textit{b,c}} were performed (Fig. 1c). The crosslinked PNIPA shell with a thickness of ca. 50 nm (light grey) is clearly visible. Fig. 1c shows that the shell is densely attached to the particle core (dark grey). Due to their statistical orientation in the vitrified water layer, the DMP particles may appear dumbbell-shaped (inset of Fig. 1c) or spherical if the main particle axis is perpendicular to the plane of the
micrograph.\\
Fig. 1d depicts a SFM phase image of the DMP particles arranged on a hydrophilic glass surface in air. The substrate, core particles and
the surrounding area can be easily distinguished in the phase image. This indicates that the PNIPA shell is spread around the core particles in accordance with Fig. 1b. Moreover, the electrostatic repulsion between the particles in solution leads to an average distance on the substrate of the order of several diameters.\\
As demonstrated in recent studies,\textsuperscript{8\textit{d},12,17} depolarized dynamic light scattering (DDLS) is a highly useful tool to study the hydrodynamics of anisometric particles. DDLS measures the translational ($D^T$) and the rotational diffusion coefficient ($D^R$). The comparison of these quantities with hydrodynamic models\textsuperscript{12} allows us to gain complementary information of the particle morphology and especially about the thickness of the crosslinked PNIPA layer in aqueous suspension. This information is essential to calculate the volume fraction of the particles in solution for different temperatures. The analysis done here follows the prescription given recently.\textsuperscript{8\textit{d},12,17} Thus, the intensity autocorrelation function is measured by DDLS. The slow relaxation rate of the intensity correlation function, $\Gamma_{\text{slow}}=D^Tq^2$, was related to translational motion, while the fast decay, $\Gamma_{\text{fast}}=D^Tq^2+6D^R$, originated from both translational and rotational motion ($q$: magnitude of the scattering vector). No coupling between rotational motion and shape fluctuations in the PNIPA network has been observed compared to previous studies with spherical core-shell microgels having a lower degree of crosslinking (2.5 mol$\%$ BIS) and a thicker PNIPA shell ($71.2\pm2$ nm at 25 $^{\rm o}$C).\textsuperscript{17}

The slow relaxation rates from DLS and DDLS experiment give the same result for $D^T$ within experimental error. The slow DDLS mode is due to a small leakage of the analyzer.\textsuperscript{12} The experimental diffusion coefficients $D^T$ and $D^R$ are shown as a function of the temperature between 14.8 and 36.8 $^{\rm o}$C in Table 1. In this range, the shrinking of the crosslinked PNIPA network leads to an increase of $D^T$ by a factor of ca. 2.6 and of $D^R$ by a factor of 5.2. This difference can be understood qualitatively since $D^T\propto R_h^{-1}$, but $D^R\propto R_h^{-3}$ within the framework of the double sphere model,\textsuperscript{13\textit{a}} where $R_h$ is the radius of one
constituent sphere.

The shell model\textsuperscript{13\textit{b}} for two interpenetrating spheres with radius $R_h$ and a center-to-center distance $l$ can be used to calculate the particle size and shape from the diffusion coefficients. This model is depicted in the inset of Fig. 2. To compare the experimental results with theoretical calculations, we proceeded as follows: from the experimental diffusion coefficients of the $\text{PMMA/PS}$ core particles without the PNIPA shell,\linebreak $D^T_{\text{core}}=(4.14\pm0.03)\times10^{-12}\text{ m}^2\text{ s}^{-1}$ and $D^R_{\text{core}}=(669\pm63)\text{ s}^{-1}$ , the values of the radius $R=42.7$ nm of the constituent spheres and the center-to-center distance $l=85.4\text{ nm}$
between the two spheres of the core were calculated using the shell model for 
$T=24.9$ $^{\rm o}$C and $\eta=0.893$ cP. Subsequently the thickness of the PNIPA shell $L_h$ (see Fig. 2) was chosen such that the theoretical diffusion coefficients $D^T_{\text{theo}}$ and $D^R_{\text{theo}}$ match best the experimental values $D^T$ and $D^R$ for each temperature assuming stick-boundary conditions. Table 1 summarizes the results. With the
exception at $T=31.8$ $^{\rm o}$C, the mean deviation between $D^T$ and $D^T_{\text{theo}}$ is less than $2\%$ and less than $3\%$ for $D^R$ and $D^R_{\text{theo}}$, respectively. The aspect ratio of the DMP in the observed temperature range is between ca. 1.4 and 1.6. Notably, the shell thickness of the DMP from an \textit{in situ} imaging technique in Fig. 1c of ca. $51.5\pm5.9$ nm is in very good agreement with the precise calculations from the hydrodynamic shell model.\\
Finally, we demonstrate that the swelling behavior of the PNIPA layer is not influenced by the geometry of the core particles. Fig. 2
shows that the values $L_h$ from Table 1 are almost identical ($\pm2$ nm) to the corresponding data obtained for a spherical reference system\textsuperscript{5\textit{a}} having the same degree of crosslinking (5 mol$\%$ BIS). Note that $L_h$ of the reference system was calculated from the translational diffusion coefficient \textit{via} the Stokes-Einstein equation and the core particle radius. This result is of particular importance for the general utility of our strategy towards tuning the shape of colloidal molecules with different morphologies.

We have reported here a strategy for the synthesis of thermoresponsive colloidal molecules with a dumbbell-shaped polymer core and a crosslinked shell of poly-(\textit{N}-isopropylacrylamide). The procedure used is simple and can be applied to core particles of any structure. The dumbbell-shaped particle morphology was directly proven \textit{in situ} by cryo-TEM. The charge induced 2-dimensional self-assembly of the novel colloids on a Si-substrate was verified by FESEM and SFM. A combined study of conventional and depolarized dynamic light scattering allowed the determination of the temperature dependent PNIPA layer thickness in the whole temperature range using the hydrodynamic shell-model. The microgel particles can change both size (200 to 300 nm) and shape (aspect ratio ca. 1.6 to 1.4) by tuning the temperature between 36.8 and 14.8 $^{\rm o}$ C. Thus, these particles are excellent model systems to study the dynamics of concentrated suspensions of non-spherical colloids.

\subsection*{References}
\begin{enumerate}
\item L. A. Lyon, Z. Meng, N. Singh, C. D. Sorrell and A. St John, 
\textit{Chem. Soc. Rev.}, 2009, \textbf{38}, 865.

\item S. Su, Md. M. Ali, C. D. M. Filipe, Y. Li and R. Pelton,
\textit{Biomacromolecules,} 2008, \textbf{9}, 935.
\item W. H. Blackburn, E. B. Dickerson, M. H. Smith, J. F. McDonald and
L. A. Lyon, \textit{Bioconjugate Chem.}, 2009, \textbf{20}, 960.
\item J.-H. Kang, J. H. Moon, S.-K. Lee, S.-G. Park, S.-G. Jang, S. Yang
and S.-M. Yang, \textit{Adv. Mater.}, 2008, \textbf{20}, 3061.
\item \textit{(a)} J. J. Crassous, M. Siebenb�rger, M. Ballauff, M. Drechsler,
O. Heinrich and M. Fuchs, \textit{J. Chem. Phys.}, 2006, \textbf{125}, 204906; \textit{(b)}
J. J. Crassous, M. Siebenb�rger, M. Ballauff, M. Drechsler,
D. Hajnal, O. Heinrich and M. Fuchs, \textit{J. Chem. Phys.}, 2008, \textbf{128},
204902.
\item M. Karg and T. Hellweg, \textit{Curr. Opin. Colloid Interface Sci.}, 2009, \textbf{14},
438.
\item \textit{(a)} J. Mrkic and B. R. Saunders, \textit{J. Colloid Interface Sci.}, 2000, \textbf{222},
75; \textit{(b)} R. Atkin, M. Bradley and B. Vincent, \textit{Soft Matter,} 2005, \textbf{1}, 160.
\item \textit{(a)} V. N. Manoharan, M. T. Elsesser and D. J. Pine, \textit{Science}, 2003,
\textbf{301}, 483; \textit{(b)} Y.-S. Cho, G.-R. Yi, S.-H. Kim, S.-J. Jeon,
M. T. Elsesser, H. K. Yu, S.-M. Yang and D. J. Pine, \textit{Chem.
Mater.}, 2007, \textbf{19}, 3183; \textit{(c)} C. S. Wagner, Y. Lu and A. Wittemann,
\textit{Langmuir}, 2008, \textbf{24}, 12126; \textit{(d)} M. Hoffmann, C. S. Wagner,
L. Harnau and A. Wittemann, \textit{ACS Nano}, 2009, \textbf{3}, 3326.
\item \textit{(a)} E. B. Mock, H. D. Bruyn, B. S. Hawkett, R. G. Gilbert and
C. F. Zukoski, \textit{Langmuir}, 2006, \textbf{22}, 4037; \textit{(b)} J.-W. Kim,
R. L. Larsen and D. A. Weitz, \textit{Adv. Mater.}, 2007, \textbf{19}, 2005.
\item \textit{(a)} D. J. Kraft, W. S. Vlug, C. M. van Kats, A. van Blaaderen,
A. Imhof and W. K. Kegel, \textit{J. Am. Chem. Soc.}, 2009, \textbf{131}, 1182; \textit{(b)}
D. J. Kraft, J. Groenewold and W. K. Kegel, \textit{Soft Matter}, 2009, \textbf{5},
3823; \textit{(c)} A. Perro, E. Duguet, O. Lambert, J.-C. Taveau,
E. Bourgeat-Lami and S. Ravaine, \textit{Angew. Chem., Int. Ed.}, 2009,
\textbf{48}, 361; \textit{(d}) C. E. Snyder, M. Ong and D. Velegol, \textit{Soft Matter},
2009, \textbf{5}, 1263.
\item A. van Blaaderen, \textit{Science}, 2003, \textbf{301}, 470.
\item M. Hoffmann, Y. Lu, M. Schrinner, M. Ballauff and L. Harnau, \textit{J.
Phys. Chem. B}, 2008, \textbf{112}, 14843.
\item \textit{(a)} B. Carrasco and J. Garcia de la Torre, \textit{Biophys. J.}, 1999, \textbf{76}, 3044;
\textit{(b)} J. Garcia de la Torre, G. Del Rio Echenique and A. Ortega, \textit{J.
Phys. Chem. B}, 2007, \textbf{111}, 55.
\item \textit{(a)} M. Vega and P. Monson, \textit{J. Chem. Phys.}, 1997, \textbf{107}, 2696; \textit{(b)}
S. R. Williams and A. P. Philipse, \textit{Phys. Rev. E: Stat. Phys.,
Plasmas, Fluids, Relat. Interdiscip. Top.}, 2003, \textbf{67}, 051301; \textit{(c)}
R. Zhang and K. S. Schweizer, \textit{Phys. Rev. E: Stat. Phys., Plasmas,
Fluids, Relat. Interdiscip. Top.}, 2009, \textbf{80}, 011502.
\item \textit{(a)} N. Dingenouts, Ch. Norhausen and M. Ballauff, \textit{Macromolecules},
1998, \textbf{31}, 8912; \textit{(b)} M. Ballauff and Y. Lu, \textit{Polymer}, 2007, \textbf{48}, 1815; \textit{(c)}
J. J. Crassous, C. N. Rochette, A. Wittemann, M. Schrinner,
M. Ballauff and M. Drechsler, \textit{Langmuir}, 2009, \textbf{25}, 7862.
\item Y. Lu and M. Drechsler, \textit{Langmuir}, 2009, \textbf{25}, 13100.
\item S. Bolisetty, M. Hoffmann, S. Lekkala, Th. Hellweg, M. Ballauff and
L. Harnau, \textit{Macromolecules}, 2009, \textbf{42}, 1264.

\end{enumerate}

\begin{figure}[ht!]
\includegraphics[width=0.5\textwidth]{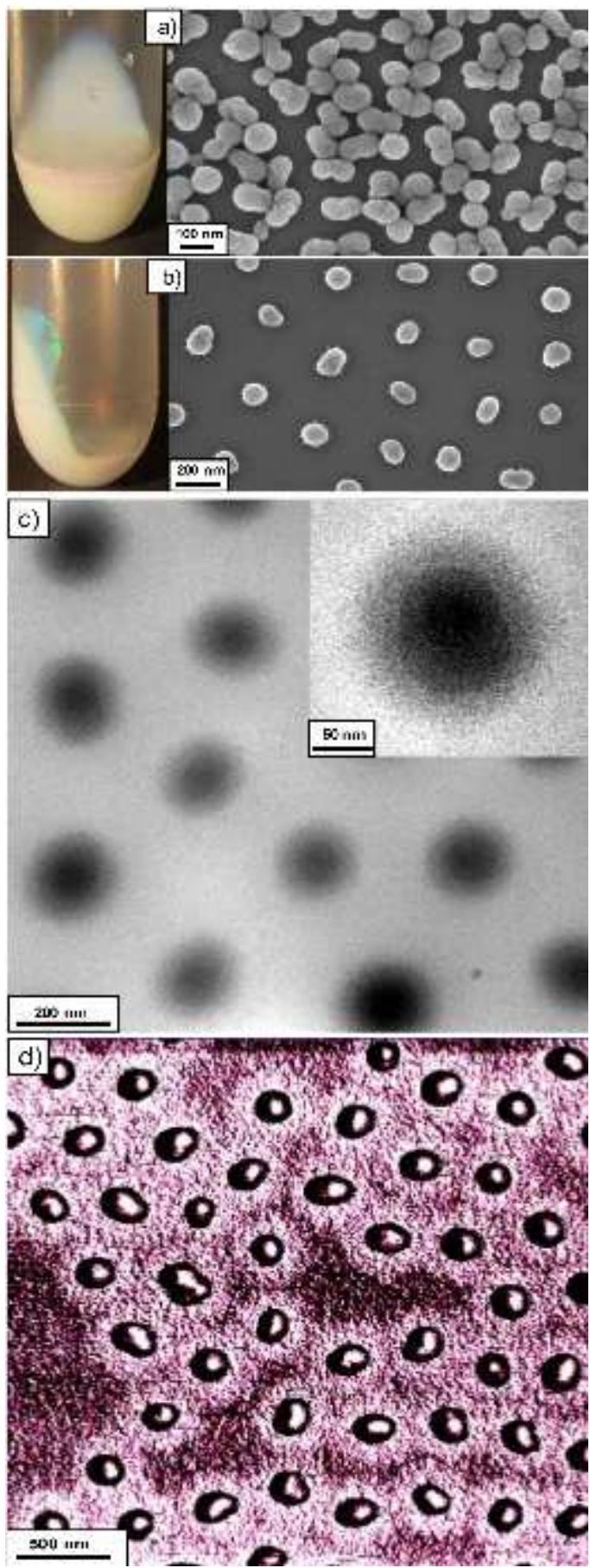}
\caption{FESEM micrographs of the a) dumbbell-shaped core particles and b) thermoresponsive core-shell particles (DMPs). The collapsed PNIPA shell spreads over a Si wafer. After centrifugation of the suspensions, iridescence can be observed. c) Cryo-TEM micrograph of the core-shell microgel particles. The shell thickness can be estimated as ${51.5\pm5.9}$ nm for $T={23}$ $^{\rm o}$C. d) SFM phase image of the DMP particles deposited from a 0.003 wt$\%$ solution on a glass slide in air (phase angle ${0-20} ^{\rm o}$).}
\label{fig:Figure1PublTCM}
\end{figure}

\clearpage
% 
%************************************************************************************ 
\begin{table}[ht!]
\begin{tabular}{cccccccc}\hline
T & $\eta$ & $D^T$ & $D^T_{\text{theo}}$ & $D^R$ & $D^R_{\text{theo}}$ &$R_h$ &$L_h$\\\hline
($^{\rm o}$C) & (cP) & ($10^{-12} {\rm m}^2{\rm s}^{-1}$)& ($10^{-12} {\rm m}^2{\rm s}^{-1}$) & (${\rm s}^{-1}$)& (${\rm s}^{-1}$)& (nm) & (nm)
\\\hline
14.8 & 1.143 & 1.41 $\pm$ 0.04 & 1.44 & 60  $\pm$ 4 & 59 & 110 & 67\\
24.5 & 0.900 & 2.04 $\pm$ 0.03 & 2.07 & 98  $\pm$ 15 & 101 & 99 & 56\\
31.8 & 0.767 & 2.80 $\pm$ 0.04 & 2.88 & 193 $\pm$ 11 & 184 & 83 & 40\\
36.8 & 0.696 & 3.71 $\pm$ 0.02 & 3.71 & 304 $\pm$ 14 & 302 & 70 & 27\\\hline
\end{tabular}
\caption{Diffusion coefficients $D^T_{\text{theo}}$ and $D^R_{\text{theo}}$ of the DMP particles as calculated with the hydrodynamic shell model together with the results for the particle size $R_h$ and $L_h$. The experimental diffusion coefficients $D^T$ and $D^R$ are also given for comparison.}
\label{tab1}
\end{table}
%************************************************************************************ 
% 

\begin{figure}[ht!]
\includegraphics[width=0.6\textwidth]{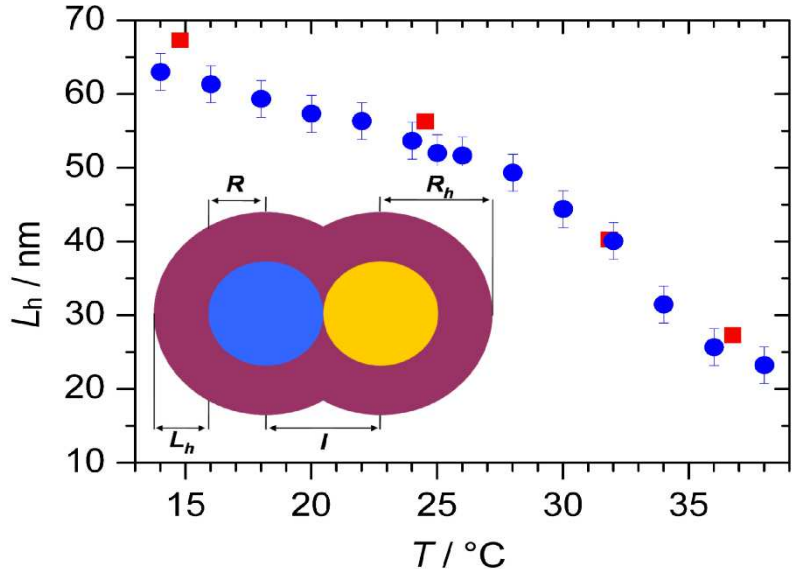}
\caption{Shell thickness $L_h$ as a function of temperature determined for the DMP particles with the hydrodynamic shell model (squares) together with the data for a spherical reference system calculated using the Stokes- Einstein equation and the core particle radius (circles). Inset: sketch of the hydrodynamic shell model used for the calculation of the shell thickness $L_h=R_h-R$. The model considers the particle surface composed of small, spherical and non-overlapping friction elements under stick-boundary conditions and allows for an interpenetration of two spheres of radius $R_h$ and a center-to-center \mbox{distance $l$.}}
	\label{fig:ModellundSchalekompakt}
\end{figure}

\end{document}